\documentclass[review]{elsarticle}

\usepackage{graphicx}%
\usepackage{amsmath,amssymb,amsfonts}%
\usepackage{amsthm}%
\usepackage{mathrsfs}%
\usepackage[title]{appendix}%
\usepackage{xcolor}%
\usepackage{textcomp}%
\usepackage{manyfoot}%
\usepackage{booktabs}%
\usepackage{algorithm}%
\usepackage{algorithmicx}%
\usepackage{algpseudocode}%
\usepackage{listings}%

\usepackage{makecell}
\usepackage{placeins}
\usepackage{multirow}
\usepackage{float}
\usepackage{graphicx}
\usepackage{graphics}
\usepackage{subcaption}
\usepackage[skins]{tcolorbox}
\usepackage{amsmath}
\usepackage{url}
\usepackage{color}
\usepackage{breqn}
\usepackage{pifont}
\usepackage{algpseudocode,caption}
\usepackage{float}
\usepackage{threeparttable}
\usepackage[inline]{enumitem}
\usepackage{multirow, booktabs}
\usepackage{hyperref}

\usepackage{xspace}
\usepackage{amsmath}
\usepackage{graphicx}
\usepackage{makecell}

\usepackage{colortbl}
\definecolor{darkgreen}{rgb}{0.0, 0.3, 0.13}
\definecolor{darkred}{rgb}{0.2, 0.0, 0.13}
\newcommand{\cc}{\cellcolor{blue!15!white}}

\newcommand{\etal}{{\em et al.}\xspace}
\usepackage{xcolor}
\newcommand{\BfPara}[1]{{\vspace{0.5em}\noindent\bf#1.}\xspace}

\usepackage{xcolor}
\usepackage{colortbl}
\usepackage{tabularx}
\colorlet{lightgrey}{lightgray}
\usepackage{tikz}
\usepackage{pgf}
\newcommand{\tsref}[1]{\textsection\ref{#1}\xspace}

\definecolor{darkgreen}{rgb}{0.0, 0.2, 0.13}
\definecolor{darkred}{rgb}{0.2, 0.0, 0.13}

\newcommand{\bad}[1]{\cellcolor{red!60!white!#1}}

\usepackage{tcolorbox}
\usepackage{colortbl}

\tcbset{textmarker/.style={%
        enhanced,
        parbox=false,boxrule=0mm,boxsep=0mm,arc=0mm,
        outer arc=0mm,left=3mm,right=3mm,top=3pt,bottom=3pt,
        toptitle=1mm,bottomtitle=1mm}}

\newtcolorbox{hintBox}{textmarker,
    borderline west={6pt}{0pt}{yellow},
    colback=yellow!10!white}

\newtcolorbox{blueBox}{textmarker,
    borderline={0pt}{0pt}{blue!50!white},
    colback=blue!10!white}
    
\newtcolorbox{importantBox}{textmarker,
    borderline west={6pt}{0pt}{red},
    colback=red!10!white}
\newtcolorbox{noteBox}{textmarker,
    borderline west={6pt}{0pt}{green},
    colback=green!10!white}

\newcommand{\observation}[1]{\begin{blueBox} \textbf{Takeaways:} #1 \end{blueBox}}

\journal{Computer Networks}

\begin{document}

\begin{frontmatter}

\title{Measuring and Modeling the Free Content Web\tnoteref{mytitlenote}}

\author{Abdulrahman Alabduljabbar$^1$, Runyu Ma$^2$, Ahmed Abusnaina$^3$, Rhongho Jang$^4$, Songqing Chen$^2$, DaeHun Nyang$^5$, and David Mohaisen$^1$}

\address{$^1$University of Central Florida, $^2$George Mason University, $^3$Meta Inc, $^4$Wayne State University, $^5$Ewha Womans University}

\tnotetext[mytitlenote]{This research was supported by Global Research Laboratory (GRL) Program through the National Research Foundation of Korea (NRF) funded by the Ministry of Science and ICT (NRF-2016K1A1A2912757). A. Alabduljabbar is supported in part by the
Saudi Arabian Cultural Mission (SACM). R. Ma and S. Chen are partly supported by a Commonwealth Cyber Initiative grant, and NSF grant CNS-2007153. (\textit{corresponding authors: David Mohaisen; \href{mailto:mohaisen@ucf.edu}{mohaisen@ucf.edu}.}). $^\eta$Equal Contributors
}

\begin{abstract}
    Free content websites that provide free books, music, games, movies, etc., have existed on the Internet for many years. While it is a common belief that such websites might be different from premium websites providing the same content types, an analysis that supports this belief is lacking in the literature. In particular, it is unclear if those websites are as safe as their premium counterparts. In this paper, we set out to investigate, by analysis and quantification, the similarities and differences between free content and premium websites, including their risk profiles. To conduct this analysis, we assembled a list of 834 free content websites offering books, games, movies, music, and software, and 728 premium websites offering content of the same type. We then contribute domain-, content-, and risk-level analysis, examining and contrasting the websites' domain names, creation times, SSL certificates, HTTP requests, page size, average load time, and content type. For risk analysis, we consider and examine the maliciousness of these websites at the website- and component-level.
Among other interesting findings, we show that free content websites tend to be vastly distributed across the TLDs and exhibit more dynamics with an upward trend for newly registered domains. Moreover, the free content websites are 4.5 times more likely to utilize an expired certificate, 19 times more likely to be malicious at the website level, and 2.64 times more likely to be malicious at the component level. Encouraged by the clear differences between the two types of websites, we explore the automation and generalization of the risk modeling of the free content risky websites, showing that a simple machine learning-based technique can produce 86.81\% accuracy in identifying them.
\end{abstract}


\end{frontmatter}

\section{Introduction}\label{sec:introduction}

Websites are categorized into two broad categories based on their monetization options: free content and premium websites. While the free content websites provide content free of charge and are typically sustained by proceeds of advertisements and user donations~\cite{Hecker99,greenhill2012,carvajal2012,Snijder10},  the premium websites offer services through fees, e.g., subscriptions or pay-as-you-use models~\cite{MayerM12}. Premium websites ensure a very high level of quality of service as a result of well-designed websites that are well-maintained through dedicated engineering and operational efforts~\cite{liang2009study}. In contrast, free content websites are believed to lack such a high expectation for the quality of service and are often user-driven~\cite{lin2013customer}. 

The lax expectations for functional and security qualities, user-driven content, and the extensive utilization of third-party advertisements on free content platforms introduce various risks~\cite{Estrada-Jimenez17,AlabduljabbarNDSS22,AlabduljabbarMC22,Alabduljabbar2022SocialNLP}. For example, advertisements on these websites can be exploited for data and information leakage or even the distribution and execution of malicious scripts on the user device~\cite{li2012knowing,GroefDNP12}. Moreover, the lack of strict maintenance operation rules in free content websites allows for various risks: web frameworks used in free content websites are rarely updated, allowing for the exploitation of old unpatched vulnerabilities and exposing their users to various levels of risk~\cite{NappaJBCD15}. 

However, untested hypotheses and widely unverified beliefs aside, are free content websites different from premium websites delivering the same type of content? Do free content websites differ in their structure, content, and security properties from premium websites? Do these websites come with a hidden cost to users, outweighing the perceived benefits, i.e., being free? To answer these questions, we proceed with a systematic analysis of a carefully assembled dataset that curates 834 free content websites and 728 premium websites. Our study combines both domain- and content-level analysis, coupled with security analysis across various dimensions. For the domain-level analysis, we examine the domain name system features, creation time, and SSL (Secure Sockets Layer) features as measures of intent. For the content-level analysis, we examine the HTTP (Hypertext Transfer Protocol) request, page size, loading time, and content type, all measuring website complexity. For security analysis, we examine both the website- and component-level detection and vulnerability using two major off-the-shelf tools, \textit{VirusTotal API}~\cite{VirusTotal} and \textit{Sucuri API}~\cite{Sucuri}.

Our analysis concludes that there are significant, fundamental, and intrinsic differences between free content and premium websites delivering the same type of content. Among other interesting findings, we report that free content websites are exclusively vastly distributed across TLDs (Top-level Domains), although using common SLDs (Second-level Domains). Moreover, they frequently change their domains, are likely to evade blacklisting, and are more often associated with invalid SSL certificates. Content-wise, free content websites tend to require significantly fewer HTTP requests for smaller requested page sizes, although at a penalty of significant load time due to extensively employing redirection with more script objects. Risk-wise, we found that free content websites are 19 and 2.64 times more likely to be malicious than premium websites at the page level (38\% vs. 2\%) and file level (45\% vs. 17\%), respectively. 

We leverage our insights from those analyses to generalize and extrapolate by modeling free content websites' risk. To this end, we defined risk using pure performance metrics. Moreover, we were able to group the risky websites with very high accuracy (more than 86\%). 

\BfPara{Contributions and Findings} This paper delivers in-depth comparative analyses of the free and premium websites of the same content types across various dimensions: domains, content, and security. Enabled by a feature-rich analysis, we build a machine learning-based approach to score the risk of free content websites with high accuracy. In the following, we elaborate on our contributions.

\begin{enumerate}
\item {\bf Free Content Websites Curation~(\tsref{sec:dataset}).} We assembled a list of more than 1,500 free content and premium websites offering the same type of content. The websites are obtained from the top search results of Google, DuckDuckGo, and Bing search engines. The websites are then crawled to obtain their content, including scripts, images, HTML, CSS, etc. 
\item {\bf Domain-level Analysis (\tsref{sec:websiteanalysis}).} To examine the domain-level features of free content websites, we analyze three aspects: their TLD (Top-level Domain), SSL certificates, and creation date. As a result, we found a significant increase in the number of free content websites, in contrast to a decrease in newly created premium websites. Moreover, we observe more frequent domain name dynamics in free content websites than in premium websites. Almost one-third of the free content websites operated using an invalid or unmatched SSL certificate.
\item {\bf Content-level Analysis (\tsref{sec:contentanalysis}).} To examine the content-level features, we analyze three aspects: the HTTP requests, page size, and average load time. Among other findings, we observe that the premium websites contain significantly more images, and their average size is three times the size of free content websites. Interestingly, however, we found the load time appears comparable due to various intrinsic design choices, including the utilization of scripts and redirection to deliver advertisements, which are more prevalent in free content websites.
\item {\bf Free Content Websites Risk Analysis~(\tsref{sec:riskAssessment}).} We leverage two popular off-the-shelf tools, \textit{VirusTotal} and \textit{Sucuri}, to assess the security risks associated with free content websites. Our analysis shows that free content websites are significantly more likely to be associated with maliciousness than premium websites. However, the discovery of premium websites detected as malicious is quite interesting and calls for further exploration.   
\item {\bf Risk Modeling (\tsref{sec:riskCharacterization}).} Both the performance and security metrics analysis highlight significant differences between free content and premium websites. Moreover, their risk profiles are vastly different from one another. Motivated by the differences in their features, we build a simple machine learning algorithm that utilizes easy-to-obtain domain- and content-level features to predict the risk of a website. We report a promising accuracy of 86.81\% for modeling the risk of free content websites. 
\end{enumerate}


\section{Related Work}\label{sec:relatedwork}
The work most related to our contribution in this paper falls under two broad branches: website analysis and malicious web content analysis. In the following, we provide an overview of some of the efforts in both directions. 

\subsection{Websites Analysis}
Websites are continuously evolving in content and usage, paralleled by an increase in the complexity and richness of their components. However, with such an evolution, various security risks emerge due to the interplay between such components~\cite{PerdicesVGP23,ZouSWCZ22,SunZCC22,FaroughiMVFMJ21}. One of the vastly unexplored security aspects in the literature has been the validity of websites' certificate~\cite{ChungLCLMMW16}. 
To address this issue, Chung~\etal~\cite{ChungLCLMMW16} proposed an in-depth analysis of certificates in the web PKI (Public Key Infrastructure), showing that the vast majority of certificates in the web PKI are invalid. 
Their study also investigated the source of the invalid certificates, concluding that they were generated mainly by end-user devices, with periodic regeneration of new self-signed certificates.

Libert~\etal~\cite{Libert15} evaluated the privacy-compromising practices employed by a million popular websites, e.g., data leakage. They concluded that roughly nine out of ten websites shared user data with third-party services without user consent. 
Using a similar dataset, Lavrenovs~\etal~\cite{LavrenovsM18} conducted a comprehensive assessment of the security of Alexa top-million websites, showing that 29.1\% of HTTPS requests have incorrect TLS (Transport Layer Security) configurations, and the HTTP Strict Transport Security (HSTS) policy is implemented in only 17.5\% of the websites. These findings are alarming and demonstrate the worrisome state of the security policies followed by such popular websites.

Exploring environments to evaluate the security flaw in web applications, Alsmadi~\etal~\cite{alsmadi18} designed a component-based testing mechanism for various invalid inputs and used this mechanism to investigate websites' behavior, including security, due to such inputs. Since the invalid input is a consistent part of the attack surface, the security of the online services and web applications is strengthened by eliminating those inputs (i.e., rejecting invalid inputs). To do so, they proposed several methods for detecting invalid inputs, uncovering many SQL injection vulnerabilities. 


\subsection{Malicious Web Content}
Recent studies have shown that adversaries are capable of embedding malicious codes within \textit{JavaScript}, \textit{GIF}, or \textit{Redictection} components of the websites~\cite{TanZLLZD18,masri2017,Patil17,ShenWJ18,WangZTZ17}. The security (and safety) of end-users depend significantly on detecting and preventing such malicious content, which has also been studied. To do so, researchers have leveraged various features of web applications, including URL (Uniform Resource Locator) domain components, webpage content, HTTP headers, and loaded scripts, and used them to detect malicious web applications~\cite{JohnsonKBD20,desai2017malicious}. It has also been shown that a promising feature set is the HTTP header information~\cite{Libert15}, where  McGahagan~\etal\cite{McGahaganBGC19} leveraged 672 of those features to build a system for malicious website detection. 
To examine the feasibility of using components and content (i.e., files and scripts) as features for detection, the authors conducted a comprehensive evaluation of different webpage content features. These 17 engineered new features can improve malicious websites' detection performance.

One crucial yet unexplored aspect of websites is the interplay between advertisements deployed on them and their associated maliciousness.
Li~\etal~\cite{LiZXYW12} investigated various malicious online advertising and marketing methods, e.g., malware propagation~\cite{MohaisenA13,KangJMK15,kang2015detecting,MohaisenAM15,AlasmaryKAPCAAN19}, click fraud, etc. Their study used a large-scale dataset of ads-related web traces, showing malicious advertisement practices in hundreds of high-ranked websites.
To examine the effectiveness of malicious advertisement detection, Masri~\etal~\cite{masri17} evaluated three tools, \textit{VirusTotal}, \textit{URLVoid}, and \textit{TrendMicro}, showing  URLVoid to provide the best performance.

Another prominent threat that has been explored is the distribution of malicious content on free download portals~\cite{GeniolaAA17}. Such portals can be maliciously utilized for distributing harmful software to end-user devices. Rivera~\etal~\cite{RiveraKSC19} conducted a systematic analysis of PUP (Potentially Unwanted Programs) and malware obtained using free download portals, showing that, on average, 8\% to 26\% of the downloaded content are either PUP or malicious.

Machine learning algorithms have also been widely used for effectively detecting malicious websites~\cite{manjeri19}. However, they are impaired by two key challenges, feature selection and evasion. To address the feature selection problem, Singh and Goyal argued for coupling the feature selection with overhead performance and accuracy in their analysis~\cite{SinghG19}. Detection evasion, the other issue, is often associated with intrinsic features, including the usage of redirection and hidden iFrames. In this domain, Liu and Lee~\cite{LiuL20} proposed an effective Convolutional Neural Network-based malicious content detection based on a screenshot of a webpage.

\BfPara{This Work} In this work, we explore and assess the maliciousness of free content websites in contrast with premium websites. Our findings show worrisome increasing trends in the portion of malicious content within free content websites. To proactively address this concern, we model the risks associated with these websites through easy-to-obtain performance features and identify up to 86.81\% of the risky websites verified against ground truth.

\begin{table*}[t]
\centering
\caption{An overview of the collected dataset. The collected URLs are associated with five different categories and belong to free content and premium websites. Overall, 1,562 websites were crawled for the purpose of this study.}
\label{tab:URLs_Files_Statistics}
\scalebox{1}{
\begin{tabular}{lrrrrrr}
\Xhline{2\arrayrulewidth}
\multirow{2}{4em}{Category}& \multicolumn{3}{c}{Free Content Websites} & \multicolumn{3}{c}{Premium Websites} \\
& URLs   & Files   & Avg. Files  & URLs    & Files    & Avg. Files\\
\Xhline{2\arrayrulewidth}
\cc{Books}    & \cc{154} & \cc{7,073}  & \cc{45.93} & \cc{195} & \cc{17,840} & \cc{91.49}\\
Games    & 80  & 6,439  & 80.49 & 113 & 11,314 & 100.12\\
\cc{Movies}   & \cc{331} & \cc{9,821}  & \cc{29.67} & \cc{152} & \cc{10,738} & \cc{70.64}\\
Music    & 83  & 6,059  & 73.00 & 86  & 7,225  & 84.01\\
\cc{Software} & \cc{186} & \cc{11,561} & \cc{62.16} & \cc{182} & \cc{18,742} & \cc{102.98}\\
\Xhline{2\arrayrulewidth}
Overall  & 834 & 40,953 & 49.10 & 728 & 65,859 & 90.47\\
\Xhline{2\arrayrulewidth}
\end{tabular}}
\end{table*}

\begin{figure*}[t]
\centering
\includegraphics[width=1\textwidth]{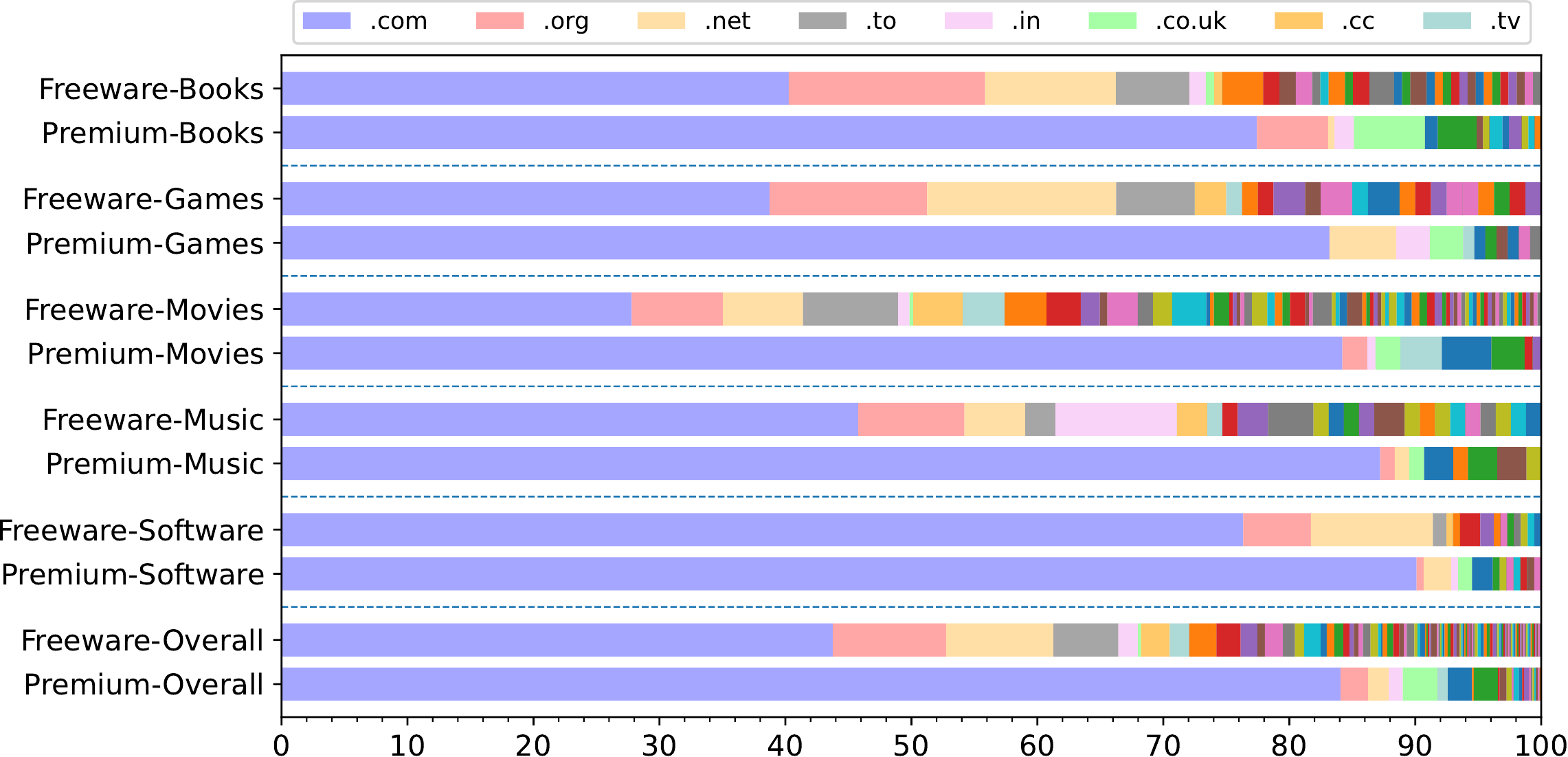}
\caption{TLD distribution of free vs. premium websites. The free content websites are more distributed among the TLDs, in contrast to the premium websites.}
\label{fig:Top_level_domain}
\end{figure*}

\section{Dataset Overview}\label{sec:dataset}
In the following, we highlight the approach we followed in creating our dataset, including initial selection and associated criteria, manual annotation, crawling, and augmentation.

\BfPara{Websites Selection} We compiled a list of 1,562 free content and premium websites for conducting our measurements. In selecting the websites, various constraints for representation. In particular, the following criteria are utilized in selecting our websites:

\vspace{0.2em}
\noindent \textbf{(1)}~\textit{\bf Popularity:} Each website has to be among the most popular websites on the web. Given that those websites may not necessarily in the most popular websites, we use search engines' results as a proxy for estimating their popularity. A website is considered popular if it appears in the top results by at least one of the used search engines: Google, DuckDuckGo, and Bing. 

\vspace{0.2em}
\noindent \textbf{(2)}~\textit{\bf Balanced Representation:} In composing our overall dataset, we ensure that our dataset is balanced per category. To that end, we expand our queries until we achieve close-to-balanced representation across categories for both the free and premium websites.  

Upon initially selecting the unique websites for inclusion in our dataset, we proceed by manually examining and labeling each of them as either premium or free content websites. 
Each of the websites is then categorized, also manually, into one of five groups based on the type of content the website mainly provides: books, games, movies, music, or software.

\BfPara{Websites Crawling} To understand the risks associated with free content websites, we crawled each website's content (i.e., files) using PyWebCopy~\cite{pywebcopy}, a python package for cloning websites and downloading their associated files. The obtained files are then used for the risk analysis and modeling, as they reflect the behavior of the provided services. Our dataset is then augmented with various attributes categorized into two broad groups, the domain-level attributes (TLD, domain creation information, SSL certificate information) and content-level attributes (HTTP request information, page size, load time, and content type).

\BfPara{High-level Characteristics} \autoref{tab:URLs_Files_Statistics} shows the distribution of the collected dataset. Notice that the average files crawled from premium websites are significantly larger than the average files for free content websites.

\section{Websites Analyses}\label{sec:analysis}
In order to understand the fundamental differences between free content and premium websites, we conduct two types of analyses: domain-level analysis and content-level analysis. Domains are the gateways to websites, and they are rich in information that can be utilized to understand their intent. Supplementing the domain-level features with content-level features improves the visibility into the websites intent. 
In the following, we provide our analysis results based on both of those features groups.

\subsection{Domain-level Analyses}\label{sec:websiteanalysis}
The domain-level analysis provides us with a high-level and interesting view and understanding of the website as an infrastructure across the owner information, creation date, and the used TLD.
We pursue such an analysis to contrast the associated domains of free and premium websites.

\BfPara{Top-level Domains Analysis}
The TLD is the highest level domains in the hierarchical domain name system, followed by the SLD (Second-level Domain); in {\tt example.com}, {\tt example} is the SLD, and {\tt com} is the TLD.
Recently, the number of TLDs has grown significantly with the introduction of the new generic TLDs (gTLDs), although \textit{`.com'}, \textit{`.net'}, \textit{`.org'}, and \textit{`.edu'} remain the most prominent~\cite{domainstate}. 
In this work, we investigate the distribution of free content and premium websites among the TLDs, shown in \autoref{fig:Top_level_domain}.

We found that \textit{`.com'} is the most prominent TLD domain, with 44\% and 84\% of free content and premium websites using  \textit{`.com'}, respectively. 
However, interesting, we found that the total number of unique TLDs used by the premium websites in our dataset to be only 24, while this number is 98 domains in the free content websites.
We note that this widespread distribution could be triggered by the mechanisms employed for malicious website blocking implemented by major browsers and systems. 
For instance, Chrome and Firefox rely on user reports when using safe browsing service~\cite{safebrowsing} to collect and block malicious websites. 
To evade blocking, free content websites change their domain name periodically. However, free content operators maintain the same SLD and migrate their websites to other TLDs to retain the existing users and some of them change their TLD to evade blocking.

\begin{figure*}[t]
    \centering
    \begin{subfigure}[t]{0.33\textwidth}
        \centering
        \includegraphics[width=0.99\textwidth]{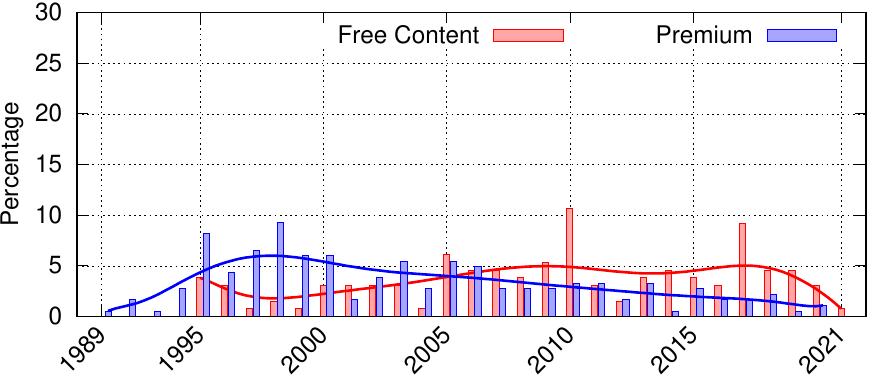}
        \caption{Books}
        \label{fig:Domain_Creation_Date_Books}
    \end{subfigure}%
    \begin{subfigure}[t]{0.33\textwidth}
        \centering
        \includegraphics[width=0.99\textwidth]{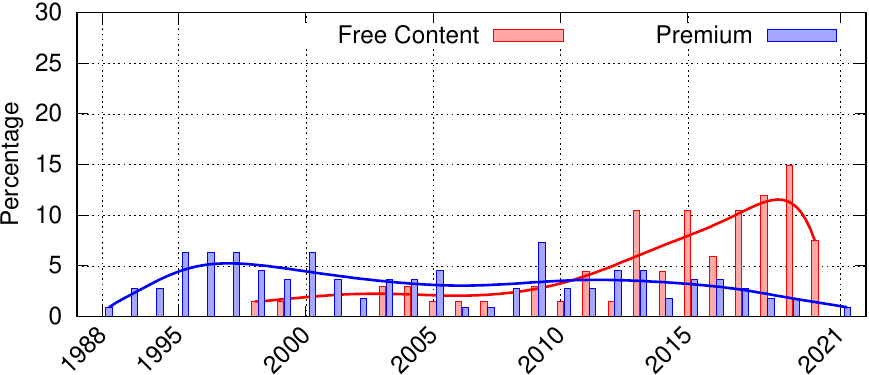}
        \caption{Games}
        \label{fig:Domain_Creation_Date_Games}
    \end{subfigure}%
    \begin{subfigure}[t]{0.33\textwidth}
        \centering
        \includegraphics[width=0.99\textwidth]{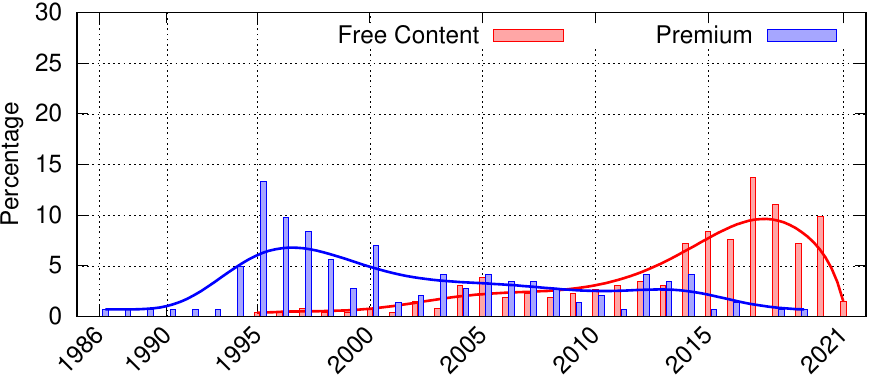}
        \caption{Movies}
        \label{fig:Domain_Creation_Date_Movies}
    \end{subfigure}
    
    \begin{subfigure}[t]{0.33\textwidth}
        \centering
        \includegraphics[width=0.99\textwidth]{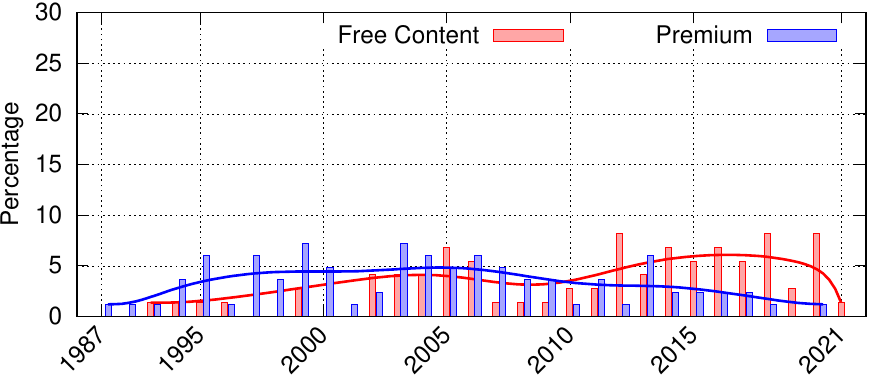}
        \caption{Music}
        \label{fig:Domain_Creation_Date_Music}
    \end{subfigure}%
    \begin{subfigure}[t]{0.33\textwidth}
        \centering
        \includegraphics[width=0.99\textwidth]{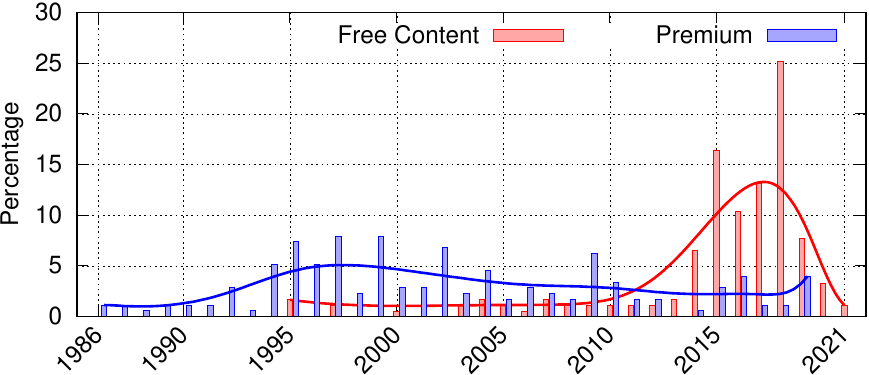}
        \caption{Software}
        \label{fig:Domain_Creation_Date_Software}
    \end{subfigure}%
    \begin{subfigure}[t]{0.33\textwidth}
        \centering
        \includegraphics[width=0.99\textwidth]{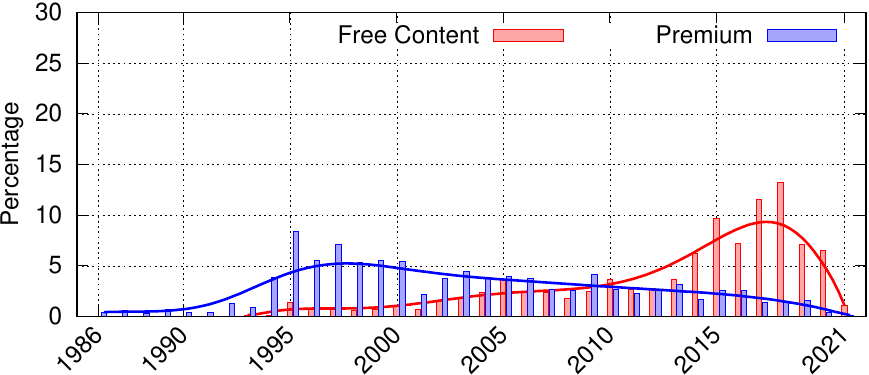}
        \caption{Overall}
        \label{fig:Domain_Creation_Date_Overall}
    \end{subfigure}
    \caption{The domain creation year comparison between free and premium website. By comparing the trend across the various content types, we observe the significant upwards trend of free content domain creation compared to premium websites.}
    \label{fig:Domain_Creation_Date}
\end{figure*}

\begin{figure}[t]
    \centering
    \begin{subfigure}[t]{0.48\textwidth}
        \centering
        \includegraphics[width=0.99\textwidth]{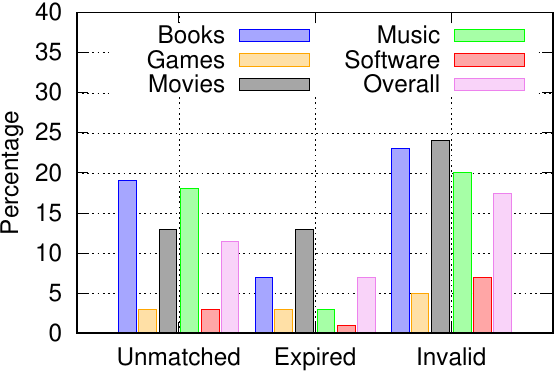}
        \caption{Free content websites}
        \label{fig:SSL_Info_Freeware}
    \end{subfigure}
    ~
    \begin{subfigure}[t]{0.48\textwidth}
        \centering
        \includegraphics[width=0.99\textwidth]{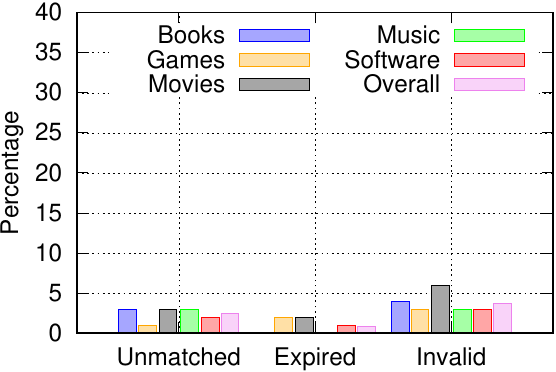}
        \caption{Premium websites}
        \label{fig:SSL_Info_Premium}
    \end{subfigure}
    \caption{The SSL certificate analysis results. We observe that almost 36\% of the free content websites have problematic SSL certificates compared to 7\% in premium websites.}
    \label{fig:SSL_Info}
\end{figure}

\begin{figure*}[t]
    \centering
        \begin{subfigure}[t]{0.31\textwidth}
        \centering
        \includegraphics[width=0.99\textwidth]{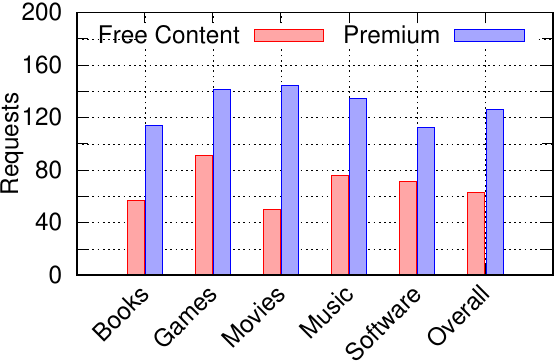}
        \caption{HTTP requests per page.}
        \label{fig:Number_of_Requests}
    \end{subfigure}%
    ~ 
    \begin{subfigure}[t]{0.31\textwidth}
        \centering
        \includegraphics[width=0.99\textwidth]{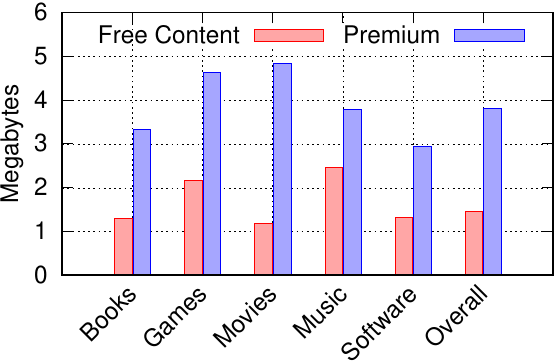}
        \caption{Size in MB.}
        \label{fig:Page_Size}
    \end{subfigure}
    ~ 
        \begin{subfigure}[t]{0.31\textwidth}
        \centering
        \includegraphics[width=0.99\textwidth]{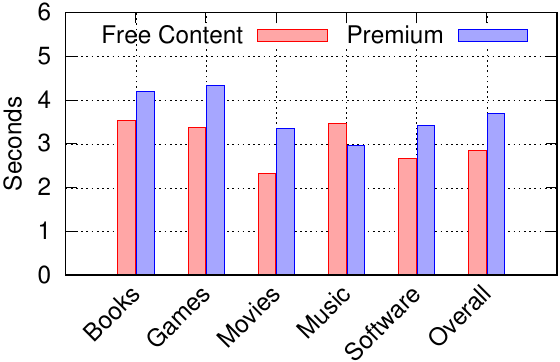}
        \caption{Load time in seconds.}
        \label{fig:Load_Time}
    \end{subfigure}
    \caption{Page-related comparison between the free content and premium websites (average statistics). Despite having different page sizes, the free content and premium websites average comparable page load times, indicating other reasons than size that affect time.}
    \label{fig:Page_Size_Load_Time_Number_of_Requests}
\end{figure*}

\BfPara{Domain Name Creation}
We examine the website creation dates, where we observe an increasing trend in the number of newly created free content websites, in contrast to the declining number of newly created premium websites, as shown in~\autoref{fig:Domain_Creation_Date}. This growing trend, particularly in the period of 2015--2021, motivates us to examine and understand the risks associated with using online free content websites. To further support that, we found from the TLDs analysis that free content websites tend to change their domain name periodically to avoid content blocking or blacklisting.

\BfPara{SSL Certificate Analysis}
HTTP transfers website content, e.g., HTML, from the web server to the user browser. However, this protocol is not secure, and the transferred data can be exposed to unauthorized access. Therefore, most websites have moved to use the secure version of HTTP (HTTPS), which implements an encryption mechanism to protect the transferred content. Our analysis found that  36\% of the free content websites have invalid HTTPS compared to only 7\% of the premium websites. Moreover, we found that 26\% of free content websites still allow HTTP (insecure) access, whereas 0\% of the premium websites allow HTTP access.
SSL certificate is a digital authentication method that authenticates the identity of a website and provides HTTPS with an encrypted connection between a server and a client machine. The SSL certificate is a critical component of a website to secure user data and protect them against, e.g., phishing.

In this work, we investigate the validity of the SSL certificate for both free content and premium websites. In particular, we study three aspects: (i) unmatched hostname in the certificate, (ii) expired certificate, and (iii) invalid/fabricated certificate.
\autoref{fig:SSL_Info} shows that, in total, 36\% of the free content websites have issues with their certificates (i.e., 11.5\% unmatched name, 7\% expired, and 17.5\% invalid certificate), compared to a total of only 7\% of the premium websites with problems in their associated SSL certificates.
This is more noticeable in the \textit{``Movies''}, \textit{``Books''}, and \textit{``Music''} categories.
As shown, free content websites are more likely to have issues with their SSL certificate. This may be attributed to the fact that free content operators are not renewing the SSL certificate, unwilling to increase their operational cost. Nonetheless, this practice leads to potential risks regarding user information and data privacy.

\observation{
Through domain-level analyses, we found that (i) the free content websites are newer, and their growth has been increasing significantly in recent years, whereas the premium websites' growth is decreasing, with fewer websites introduced every year, (ii) the free content websites are more distributed across the TLDs as they change their domain to avoid malicious website blocking mechanisms, (iii) the free content websites are more likely to have invalid or expired SSL certificate. These findings complement our analysis concerning the safety of using free content websites and the risks associated with them.
}

\begin{figure}[t]
\centering
\includegraphics[width=1\textwidth]{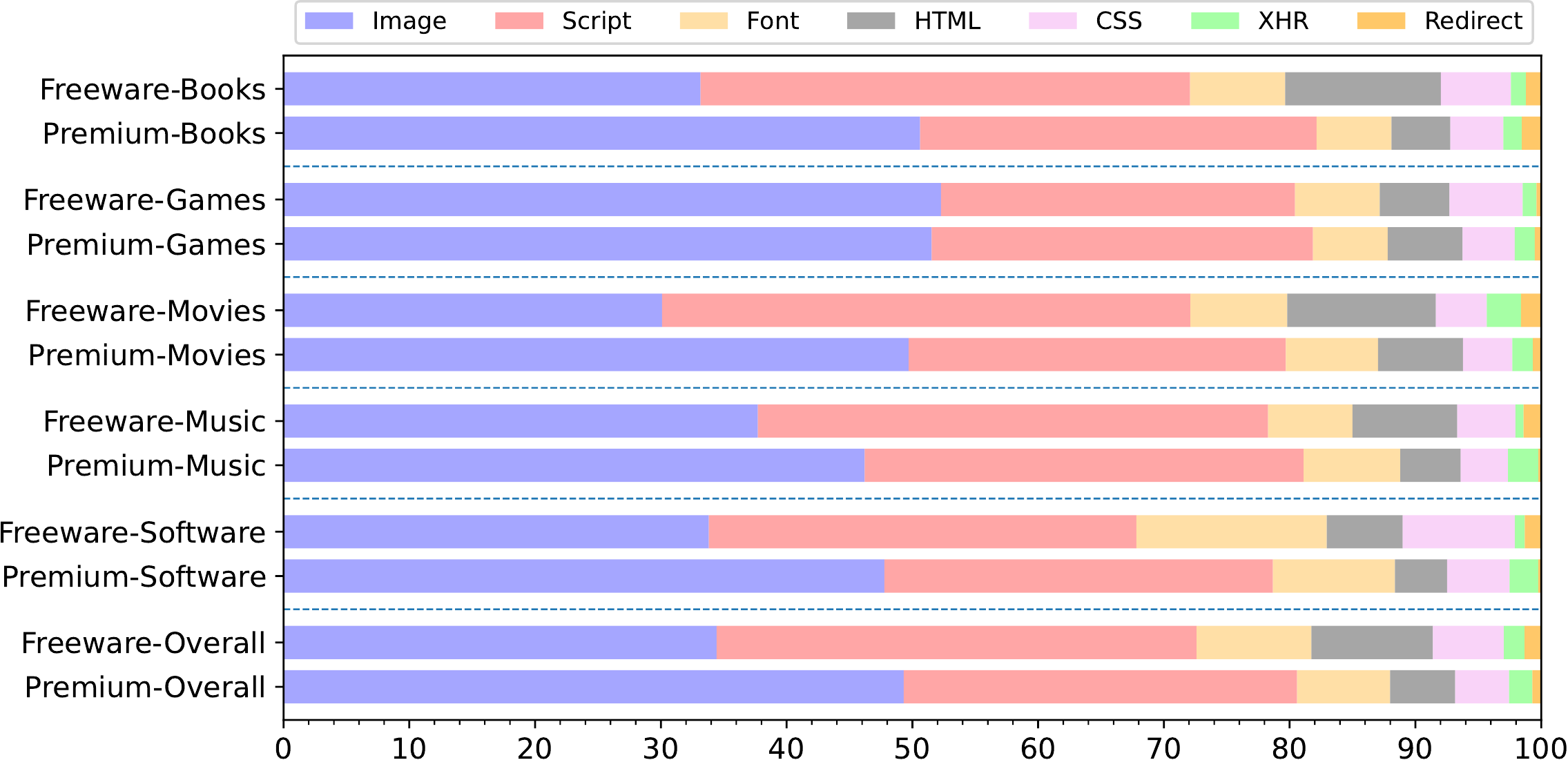}
\caption{Content-type comparison between the free content and the premium websites. We observe some differences in the website file types, notably in the {\it images} type.}
\label{fig:Content_Type}
\end{figure}

\subsection{Content-level Analyses}\label{sec:contentanalysis}
To gain insight into the content-level features of free content and premium websites, we analyze the extracted files in both types of websites. In this analysis, we focus on four features: the number of HTTP requests, page size, page load time, and content type. 

\BfPara{HTTP Requests}
HTTP requests are made by clients to request access to resources on servers (e.g., HTML files, CSS, images), and their numbers per page are an indication of the complexity of the requested page. \autoref{fig:Number_of_Requests} shows the average number of HTTP requests made for free content and premium websites. We observe that a client would initiate almost twice the number of requests to access a premium website compared to accessing a free content website. This is quite anticipated, given that the premium website pages are larger in size. However, we observe that the average page size in premium websites is 3x the free content websites, whereas the number of HTTP requests is only 2x more, indicating that visiting a free content page requires more HTTP requests for the same amount of data. That could be a result of redirection, where each redirection triggers one or more independent HTTP requests and consumes more time for loading, as shown later. 

\BfPara{Average page size} According to the page weight report by HTTP Archive~\cite{HTTPArchive}, the average page size of the top one million websites is around 2.07 MB. How far is the size of the average page that belongs to either category? To answer this question, we examine the average page size of the free content and premium websites, with the results reported in~\autoref{fig:Page_Size}. We observe that the free content websites follow the normal distribution of the page sizes reported by the HTTP Achieve~\cite{HTTPArchive}, while the premium websites have an average homepage size of 3.9MB, three times the average size of a free content page.  A potential explanation might be that the free content websites rely on redirecting users to other websites' content or advertisement websites, as we demonstrate later, instead of including and presenting content in the free content page body.

\BfPara{Average page load time}
We define page load time as the time it takes the page to be loaded fully and measured to understand additional aspects of the website's complexity.  
\autoref{fig:Load_Time} shows the average page load time, calculated using the SolarWinds Pingdom API (Application Programming Interface)~\cite{Pingdom}, for both the free content and premium websites. While the average size of the premium websites is three times the average free content page size, we notice that the average load time is comparable across them, indicating aspects beyond the size that affect the load time, i.e., degraded performance and extensive usage of redirection.

\BfPara{Content type} The page size does not seem to fully explain the complexity and loading time of websites, which calls for a deeper analysis of the content of the website. The content type is another statistical feature of the website's content at the component level (i.e., files).
These components include \textit{Image} (\textit{GIF}, \textit{PNG}, \textit{JPEG}), \textit{JavaScript}, \textit{Text}, \textit{HTML}, \textit{CSS}, \textit{XHR}, and \textit{Redirection}. We found that \textit{Image} is the most common component, followed by \textit{JavaScript}, whereas the \textit{Redirection} content is the least common among these components.
\autoref{fig:Content_Type} shows the average distribution (\%) of the different components in the free content and premium websites. Overall, premium websites have 15\% more images than free content websites.
However, we notice the extensive usage of \textit{Redirection} in the free content websites, as it is often a method to deliver advertisements and mislead the filtering algorithm. We found that the (rounded) ratio of the redirection in free content compared to premium pages to be 6 (software), 7 (music), 3 (movie), 1 (games), 1 (books). Overall, free content websites redirect twice as much as premium sites, and have twice the HTML, 1.5 times the CSS, and 1.23 times the JavaScript.

\begin{figure*}[t]
\begin{minipage}[c]{.66\textwidth}
    \centering
    \begin{subfigure}[t]{0.5\textwidth}
        \centering
        \includegraphics[width=0.99\textwidth]{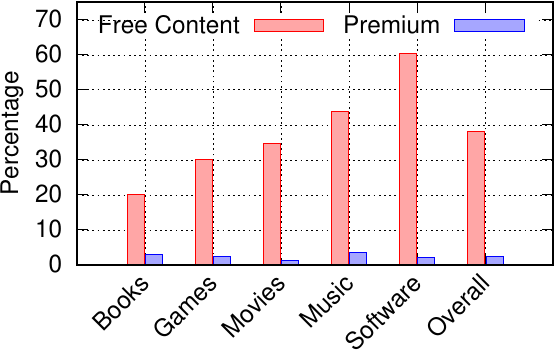}
        \caption{Labeled as malicious.}
        \label{fig:VirusTotal_URLs}
    \end{subfigure}%
    \begin{subfigure}[t]{0.5\textwidth}
        \centering
        \includegraphics[width=0.99\textwidth]{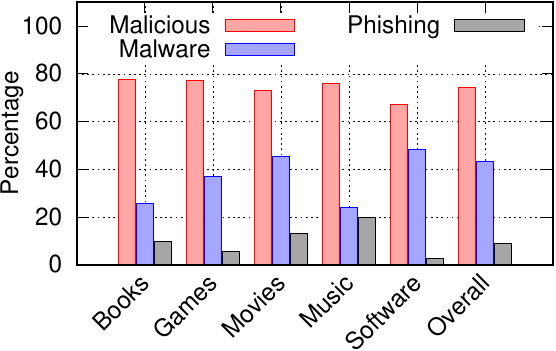}
        \caption{Malicious FCWs.}
        \label{fig:VirusTotal_Freeware_Labels}
    \end{subfigure}
    \caption{The potential maliciousness of free content and premium websites.}
    \label{fig:VirusTotal_URLs_Labels_Engines}
    \end{minipage}~
    \begin{minipage}[c]{.33\textwidth}
    \centering
\includegraphics[width=0.99\textwidth]{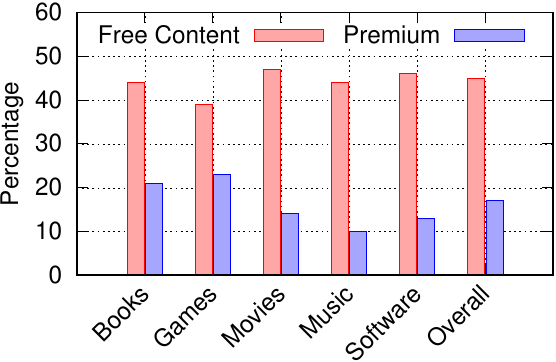}
\caption{The malicious files detected by \textit{VirusTotal}: free content vs. premium.}
\label{fig:VirusTotal_Files}
\end{minipage}
\end{figure*}

\observation{
Our content-level analyses shed light on the main differences between free content and premium websites. We found the following. (i) The premium websites have almost twice the number of requests as free content websites and three times the average size of free content websites pages. (ii) Nevertheless, the average homepage load time is comparable for free content and premium websites. (iii) Content type-wise, the free content websites have a higher portion of \textit{redirection} components, as they are a primary method to deliver advertisements.}

\section{Maliciousness Analyses}\label{sec:NFL}
The analysis we conducted so far considered the performance and non-security characteristics of free content and premium websites, which highlight clear differences that contribute to both direct and indirect costs. One of the most important and obvious metrics to measure the cost of free content websites is by understanding their security and associated risk. In this section, we conduct this analysis, focusing on indicators of threat, such as maliciousness of URLs, files, and associated vulnerabilities (\textsection\ref{sec:riskAssessment}). Towards automating the discovery of such risks, we also report the results of a machine learning-based tool that shows the risk boundaries of websites based on features obtained from the risk analysis (\textsection\ref{sec:riskCharacterization}).

\subsection{Risk Assessment}\label{sec:riskAssessment}
The study of the maliciousness and vulnerabilities of both services websites, by shedding examining how they potentially affect users experience, safety, and security, is important. Motivated by that, we define the risk of a website using several metrics, namely: (i) containing malware, (ii) running malicious scripts, (iii) exploiting user device's resources, or (iv) containing vulnerabilities, outdated software versions, or unpatched frameworks. 

To assess the risk of each type of website without reinventing the wheel, we leverage two public APIs: \textit{VirusTotal}~\cite{VirusTotal} and \textit{Sucuri}~\cite{Sucuri} for harmful behavior analysis. \textit{VirusTotal} is an online service that aggregates the scanning results of more than 70 scanning engines and can be used for scanning files and URLs alike. On the other hand, Sucuri is a service that tests websites against several known malware, viruses, blacklisting lists, vulnerabilities, outdated frameworks, and malicious code.

\BfPara{Malicious URLs Detection and Annotation}
Using \textit{VirusTotal} API, we extracted malicious activities associated with the website URL, shown in~\autoref{fig:VirusTotal_URLs_Labels_Engines}. 
We notice that there is a noticeable discrepancy between free content and premium websites in terms of maliciousness. In particular, \autoref{fig:VirusTotal_URLs} shows that 38\% of the free content websites are considered malicious by \textit{VirusTotal}, compared to only 2\% of the premium websites. A significant number of those detected websites ($\approx$74\%) were labeled as malicious (\autoref{fig:VirusTotal_Freeware_Labels}), a website created to promote scams, attacks, and frauds. We also notice that a significant portion of the free content URLs is detected as malicious, ranging from 20\% (\textit{``Books''} websites)  to 60\% (\textit{``Software''} websites). In contrast, premium websites have a very low detection rate, ranging from 1\% to 4\% only.

\begin{table*}[t]
\caption{The distribution of malicious files for different file formats in free content and premium websites. We observe that a large portion of \textit{``.gif''} files are labeled as malicious in both cases, although almost twice as much (percentage) in free content.}

\label{tab:Malicious_Files_Formats}
\scalebox{0.8}{
\begin{tabular}{clrrrrrrrrrrrrr}
\Xhline{2\arrayrulewidth}

 & Category & .gif & .html & .png & .js & .php & .woff & .jpg & .eot & .woff2 & .svg & .ttf & .log & .css \\ \Xhline{2\arrayrulewidth}

\multirow{5}{*}{\rotatebox[origin=c]{90}{Free Content}} & \cc{Books}    & \cc{\bad{56}28\%} & \bad{2}\cc{1\% }  & \cc{0\% } & \bad{4}\cc{2\%} & \cc{0\%}  & \cc{0\%}   & \cc{0\%}  & \cc{0\%}  & \cc{0\%}    & \cc{0\%}  & \cc{0\%}  & \cc{0\%}  & \cc{0\%}  \\
& Games    & \bad{14}7\%  & \bad{26}13\%  & 0\%  & \bad{2}1\% & \bad{6}3\%  & 0\%   & 0\%  & 0\%  & 0\%    & 0\%  & 0\%  & 0\%  & 0\%  \\
& \cc{Movies }  & \bad{80}40\% & \bad{12}\cc{6\%}   & \bad{2}\cc{1\%}  & \bad{4}\cc{2\%} & \cc{0\%}  & \bad{2}\cc{1\%}   & \cc{0\%}  & \bad{2}\cc{1\%}  & \cc{0\%}    & \bad{2}\cc{1\%}  & \bad{2}\cc{1\%}  & \bad{2}\cc{1\%}  & \cc{0\%}  \\
& Music    & \bad{26}26\% & \bad{6}6\%   & 0\%  & \bad{1}1\% & \bad{8}4\%  & \bad{6}3\%   & 0\%  & \bad{6}3\%  & \bad{6}3\%    & 0\%  & \bad{4}2\%  & 0\%  & 0\%  \\
& \cc{Software} & \bad{22}\cc{11\%} & \bad{60}30\%  & \bad{8}\cc{4\%}  & \bad{1}\cc{1\%} & \bad{2}\cc{1\%}  & \bad{8}\cc{4\%}   & \cc{0\%}  & \bad{4}\cc{2\%}  & \bad{8}\cc{4\%}    & \bad{4}\cc{2\%}  & \bad{6}\cc{3\%}  & \bad{2}\cc{1\%}  & \cc{0\%}  \\ 
\hline
& Overall  & \bad{52}26\% & \bad{22}11\%  & \bad{4}2\%  & \bad{1}1\% & \bad{4}2\%  & \bad{4}2\%   & 0\%  & \bad{4}2\%  & \bad{8}4\%    & \bad{2}1\%  & \bad{4}2\%  & \bad{2}1\%  & 0\%  \\
\Xhline{2\arrayrulewidth}

\multirow{5}{*}{\rotatebox[origin=c]{90}{Premium }} & \cc{Books }   & \bad{38}19\% & \bad{2}\cc{1\%}   & \cc{0\%}  & \cc{0\%} &\cc{0\%}  & \cc{0\%}   & \cc{0\%}  & \cc{0\%}  & \cc{0\%}    & \cc{0\%}  & \cc{0\%}  & \cc{0\%}  & \cc{0\%}  \\
& Games    & \bad{42}21\% & 0\%   & 0\%  & 0\% & 0\%  & 0\%   & 0\%  & 0\%  & 0\%    & 0\%  & 0\%  & 0\%  & 0\%  \\
& \cc{Movies}   & \bad{18}9\%  & \bad{2}\cc{1\%}   & \cc{0\%}  & \cc{0\%} & \cc{0\%}  & \cc{0\%}   & \cc{0\%}  & \cc{0\%}  & \cc{0\%}    & \cc{0\%}  & \cc{0\%}  & \cc{0\%}  & \cc{0\%}  \\
& Music    & \bad{42}21\% & \bad{2}1\%   & 0\%  & 0\% & 0\%  & 0\%   & 0\%  & 0\%  & 0\%    & 0\%  & 0\%  & 0\%  & 0\%  \\
& \cc{Software} & \bad{10}5\%  & \bad{2}\cc{1\%}   & \cc{0\%}  & \cc{0\%} & \cc{0\%}  & \cc{0\%}   & \cc{0\%}  & \cc{0\%}  & \cc{0\%}    & \cc{0\%}  & \cc{0\%}  & \cc{0\%}  & \cc{0\%}  \\
\hline
& Overall  & \bad{30}15\% & \bad{2}1\%   & 0\%  & 0\% & 0\%  & 0\%   & 0\%  & 0\%  & 0\%    & 0\%  & 0\%  & 0\%  & 0\% \\ \Xhline{2\arrayrulewidth}
\end{tabular}}
\end{table*}

\begin{figure}[t]
    \centering
    \begin{subfigure}[t]{0.48\textwidth}
        \centering
        \includegraphics[width=0.98\textwidth]{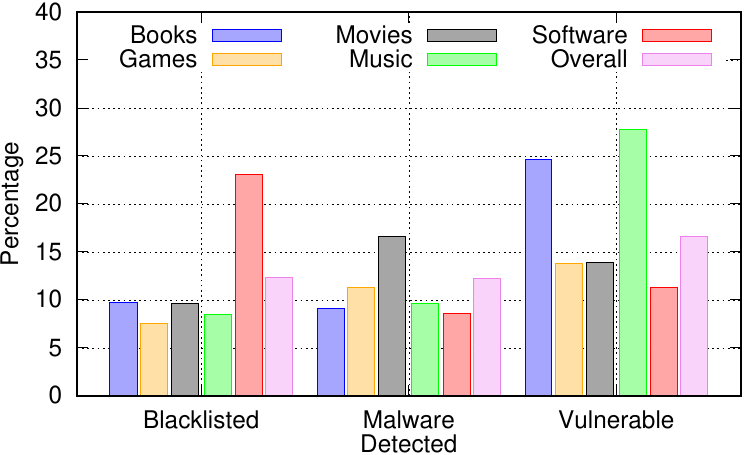}
        \caption{Free content websites}
        \label{fig:Sucuri_Freeware}
    \end{subfigure}
    ~
    \begin{subfigure}[t]{0.48\textwidth}
        \centering
        \includegraphics[width=0.98\textwidth]{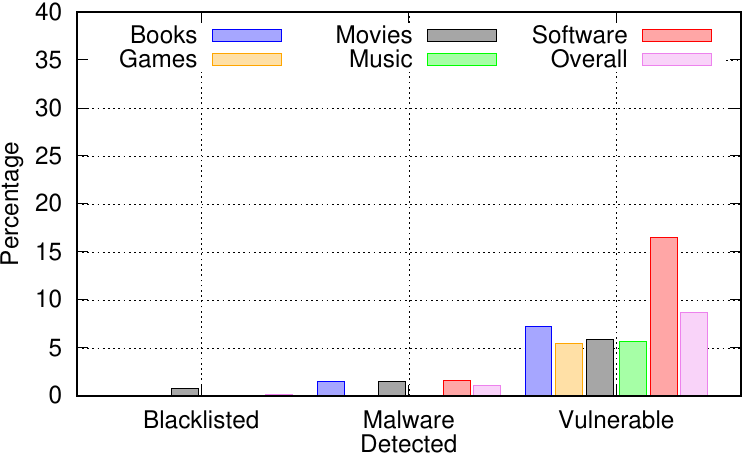}
        \caption{Premium websites}
        \label{fig:Sucuri_Premium}
    \end{subfigure}
    \caption{Assessing the maliciousness of the free content and premium websites. We show the percentage of the websites labeled as blacklisted, malware, and vulnerable.}
    \label{fig:Sucuri_Vulnerability_Blacklisting}
\end{figure}

\BfPara{Malicious File Detection and Formats Analysis}
In order to understand the behavior of a given service (i.e., content providers), it is essential to analyze the behavioral characteristics of the executable scripts hosted by the service. These scripts are forwarded to the end-user as files, including images, JavaScript codes, HTML, among other formats, and are often rendered or executed on the user's device. Analyzing the scripts and website files is critical, as recent studies~\cite{CohenNE20,SouthSB21,YostJ17,JungLE20} have shown that such content can be exploited, leading to information and data leakage, in addition to abusing the resources of the end-user device. 
In order to understand the risks of free content websites, we leverage \textit{VirusTotal} API for malicious file identification. In contrast, \autoref{fig:VirusTotal_Files} shows the percentage of malicious files detected by the \textit{VirusTotal} API in the free content and premium websites. While the number of URLs that pertain to the premium websites and are labeled as malicious is only 2\%, the number of their files labeled as malicious was 17\%.

We notice that the trend persists overall, although magnified: 45\% of the free content websites had files that have been labeled as malicious (compared to 17\% in premium). To better understand this observation, we investigate the distribution of the format of the malicious files, where the comparative results are shown in~\autoref{tab:Malicious_Files_Formats}. 

Based on \autoref{tab:Malicious_Files_Formats}, we report that the majority of malicious files have \textit{`.gif'} and \textit{`.html'} formats. This is a result of either (i) the \textit{`.gif'} and \textit{`.html'} files containing malicious embedded scripts, or (ii) the VirusTotal engines considering the \textit{`.gif'} files as malicious content in general (i.e., potential false positives). It is worth noting that we manually inspected the \textit{`.gif'} files, and found that the majority of the malicious-labeled  \textit{`.gif'} files are advertisement-related content.

\BfPara{Websites Vulnerability and Blacklisting}
In order to analyze the potential exploitable vulnerabilities and blacklisting, we leveraged Sucuri API~\cite{Sucuri} to obtain information of domains activities for both types of services. 
As a result, we found that 12\% of the free content websites were detected as \textit{containing malware}, compared to only 1\% of their premium counterparts, as shown in~\autoref{fig:Sucuri_Vulnerability_Blacklisting}. Moreover, we found the free \textit{``Movie''} websites have the highest percentage of malware detection (16.67\%), as shown in~\autoref{fig:Sucuri_Freeware}. 

We also scanned the websites for vulnerabilities and found that the free \textit{``Books''} and \textit{``Music''}  websites have the highest vulnerabilities overall. 
Despite the low reporting rate in the premium websites, 17\% of \textit{``Software''} were labeled as vulnerable, a higher portion than in free content websites (12\%), which is quite surprising. According to \textit{Sucuri} reports, a high percentage of the legitimate \textit{``Software''} websites vulnerabilities are due to outdated framework versions, which is common in \textit{``Software''} services websites.

In terms of blacklisting,~\autoref{fig:Sucuri_Freeware} shows that 12\% of the free content websites were blacklisted by the \textit{Sucuri} scanning engines, including Google, McAfee, Yandex, Norton, ESET, and AVAST engines. We observe that the \textit{``Software''} free content websites have a significantly higher percentage of blacklisted URLs (23.12\%) compared to other categories, which all had at most 12\% blacklisting rate. 
One reason for this behavior is the fact that these websites are changing their domain names frequently using a different TLD.

\observation{
To assess the risks associated with free content websites, we leveraged \textit{VirusTotal} and \textit{Sucuri} APIs for analyzing the maliciousness of domain and files of both service types. Our analyses show worrisome trends among free content websites, including (i) free content websites are more likely to be associated with maliciousness at a domain-level (38\% of the free content websites), and (ii) they are more likely to be associated with maliciousness at the file-level (45\% of them). These trends are not limited to maliciousness, which led to high blacklisting, but include exploitable vulnerabilities that can expose visitors to leakage attacks. Our analysis also unveils that 17\% of the free content websites are vulnerable. }

\begin{table*}[t]
\caption{The description of the website's characterization features. The features are extracted from three sources, (i) The website's content, (ii) The website's public information, (iii) The website's SSL certificate information. We include the characteristics extracted from VirusTotal and Sucuri APIs for risk characterization and potential detection (D). Org: the origin of the feature, \textit{[c]}: categorical feature, \textit{[b]}: boolean feature (T/F), \textit{[n]}: numerical feature, \textit{[p]}: percentage feature.}
     \label{tab:features_description}
     \scalebox{0.7}{
\begin{tabular}{lccl!{\vrule width 0.8pt}lccl}
\Xhline{2\arrayrulewidth}
\textbf{\#} & \textbf{\textsection} & Type          & \textbf{Description}                                                        & \textbf{\#} & \textbf{\textsection}  & Type                   & \textbf{Description}                                                    \\ \Xhline{2\arrayrulewidth}

\cc{1}  & \cc{~\tsref{sec:websiteanalysis}} & \cc{[c]} & \cc{TLD name used by the website }                                  & \cc{14} & \cc{~\tsref{sec:contentanalysis}} & \cc{[p]}         & \cc{\% of Redirect content in the website}              \\ 
2  & ~\tsref{sec:websiteanalysis} & [b]& Domain not matched   & 15 & ~\tsref{sec:riskAssessment} & [b]  & Domain detected by VirusToal API              \\ 
\cc{3}  & \cc{~\tsref{sec:websiteanalysis}} & \cc{[b]}& \cc{Expired SSL certificate}                      & \cc{16} & \cc{~\tsref{sec:riskAssessment}} & \cc{[b]} & \cc{Website is detected as malicious}                  \\ 
4  & ~\tsref{sec:websiteanalysis} & [b]& SSL certificate cannot be verified                    & 17 & ~\tsref{sec:riskAssessment} & [b]    & Website containing malware            \\ 
\cc{5}  & \cc{~\tsref{sec:contentanalysis}} & \cc{[n]}& \cc{Average number of HTTP requests}                    & \cc{18} & \cc{~\tsref{sec:riskAssessment}} & \cc{[b]}    & \cc{Website detected as phishing}                      \\ 
6  & ~\tsref{sec:contentanalysis} & [n]& Average content size                        & 19 & ~\tsref{sec:riskAssessment} & [b]  & Files detected as malicious  \\
\cc{7}  & \cc{~\tsref{sec:contentanalysis}} & \cc{[n]}& \cc{Page load time} & \cc{20} & \cc{~\tsref{sec:riskAssessment}} & \cc{[b]}        & \cc{URL detected as malicious}                 \\ 
8  & ~\tsref{sec:contentanalysis} & [p]& \% of images in the website& 21 & ~\tsref{sec:riskAssessment} & [b]  &Blacklisted by scanning engines           \\ 
\cc{9}  & \cc{~\tsref{sec:contentanalysis}} & \cc{[p]}   & \cc{\% of script files in the website}  & \cc{22} & \cc{~\tsref{sec:riskAssessment}} & \cc{[b]}           & \cc{Vulnerability in the website } \\ 
10 & ~\tsref{sec:contentanalysis} & [p]  & \% of Fonts content in the website   & 23 & ~\tsref{sec:riskCharacterization} & [n]       & Website's IP address lifetime     \\ 
\cc{11} & \cc{~\tsref{sec:contentanalysis}} & \cc{[p]}    & \cc{\% of HTML files in the website}& \cc{24} & \cc{~\tsref{sec:riskCharacterization}} & \cc{[b]}      & \cc{Using/used Cloudflare as a CDN}               \\ 
12 & ~\tsref{sec:contentanalysis} & [p]      & \% of CSS files in the website  & 25 & ~\tsref{sec:riskCharacterization} & [b]   & Using/used Akamai as a CDN               \\ 
\cc{13} & \cc{~\tsref{sec:contentanalysis}} & \cc{[p]}    & \cc{\% of XHR content in the website}   &   \cc{}   &   \cc{}  &\cc{}              &  \cc{}  \\  \Xhline{2\arrayrulewidth}
\end{tabular}}
\end{table*}

\begin{figure*}[thb]
    \centering
    \begin{subfigure}[t]{0.19\textwidth}
        \centering
        \includegraphics[width=0.99\textwidth]{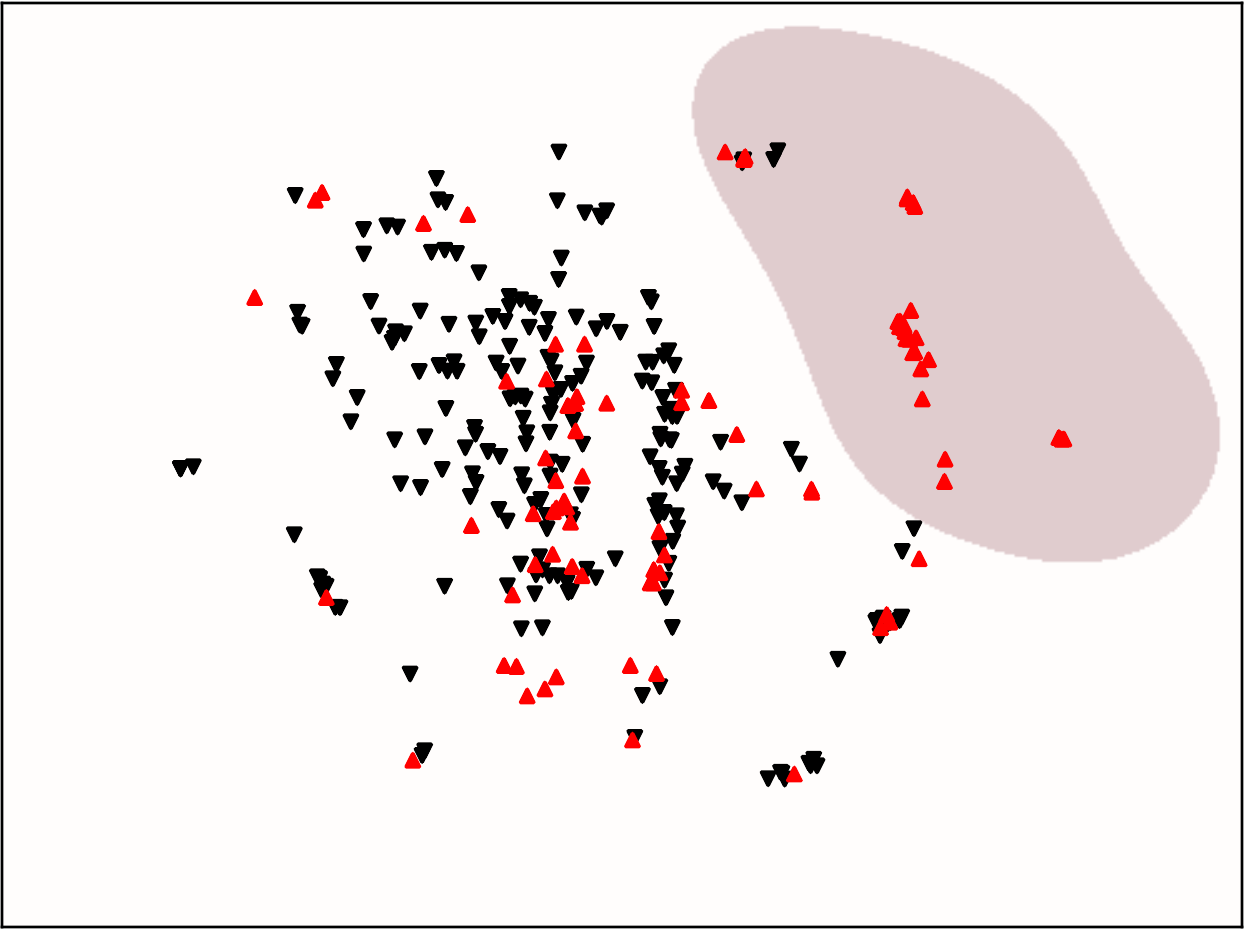}
        \caption{Books}
        \label{fig:ML_Clustering_Books2}
    \end{subfigure}~
    \begin{subfigure}[t]{0.19\textwidth}
        \centering
        \includegraphics[width=0.99\textwidth]{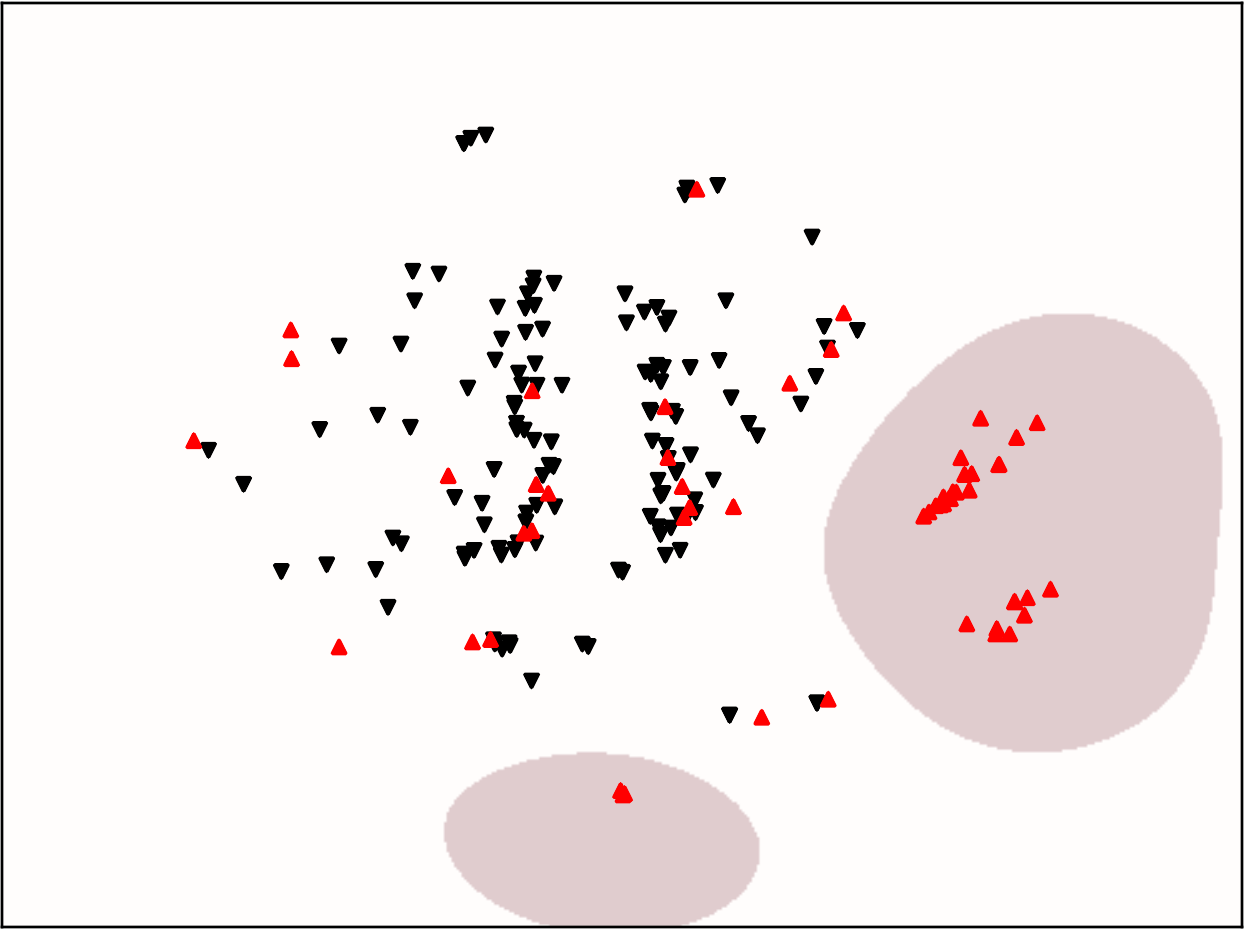}
        \caption{Games}
        \label{fig:ML_Clustering_Games2}
    \end{subfigure}~
    \begin{subfigure}[t]{0.19\textwidth}
        \centering
        \includegraphics[width=0.99\textwidth]{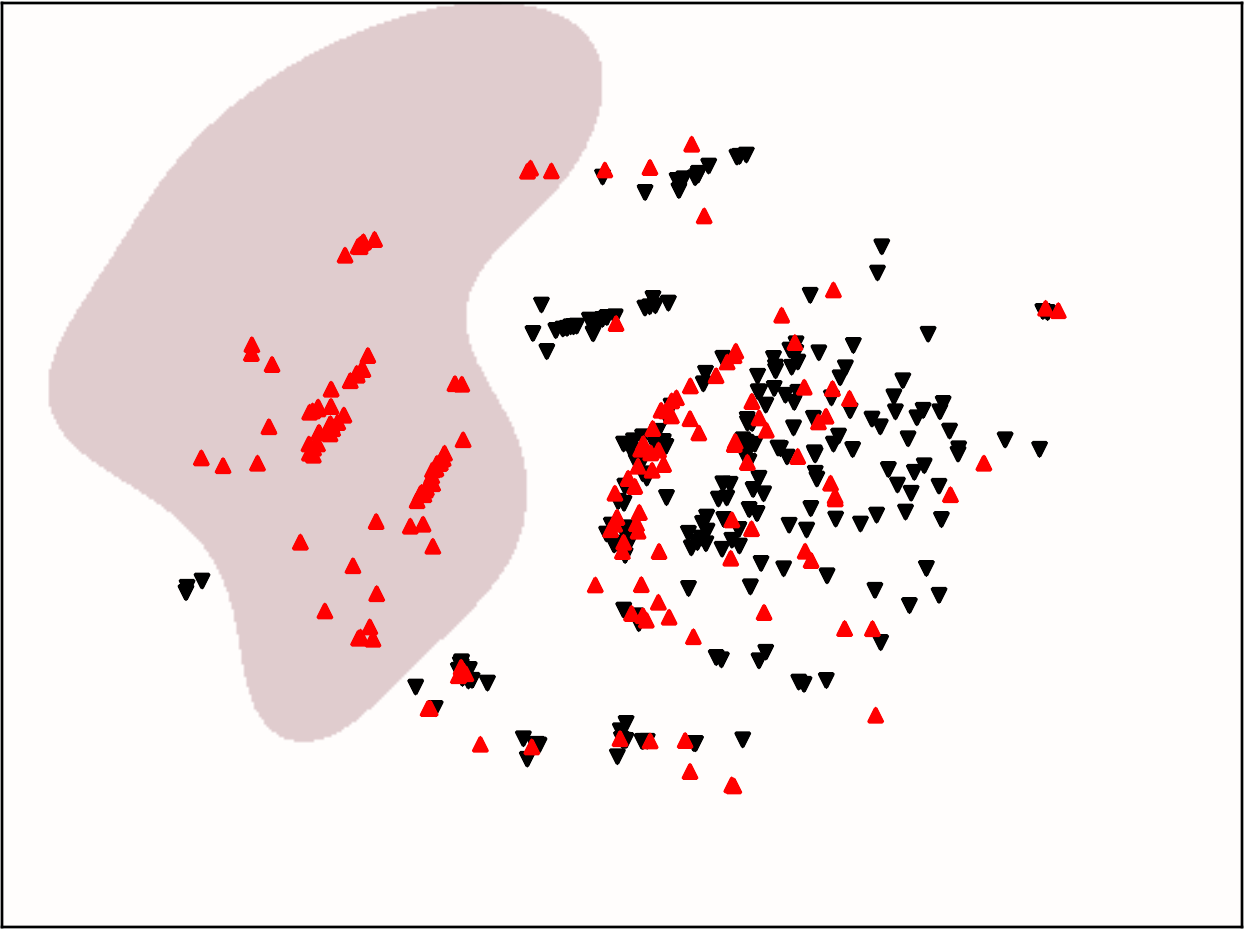}
        \caption{Movies}
        \label{fig:ML_Clustering_Movies2}
    \end{subfigure}~
    \begin{subfigure}[t]{0.19\textwidth}
        \centering
        \includegraphics[width=0.99\textwidth]{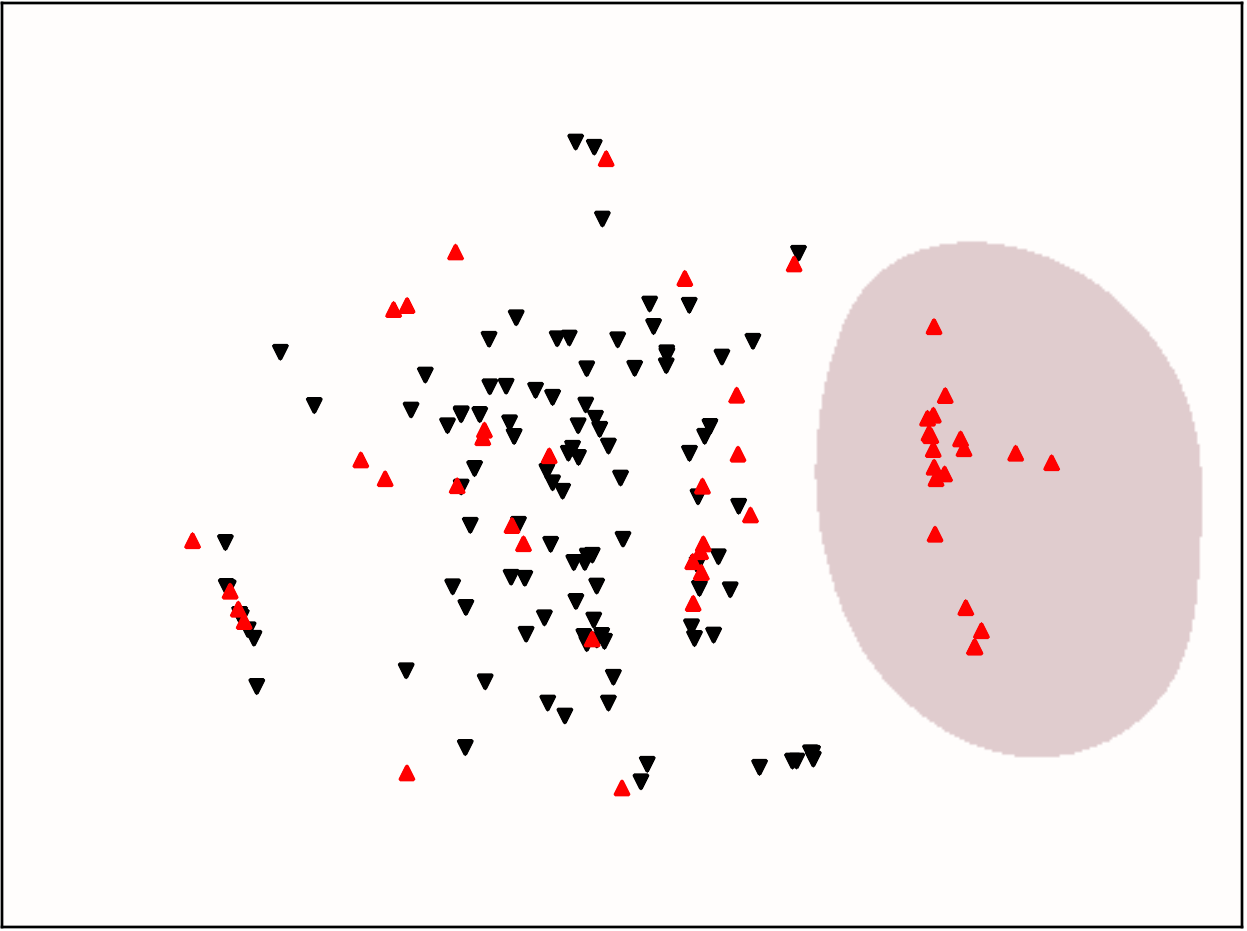}
        \caption{Music}
        \label{fig:ML_Clustering_Music2}
    \end{subfigure}~
    \begin{subfigure}[t]{0.19\textwidth}
        \centering
        \includegraphics[width=0.99\textwidth]{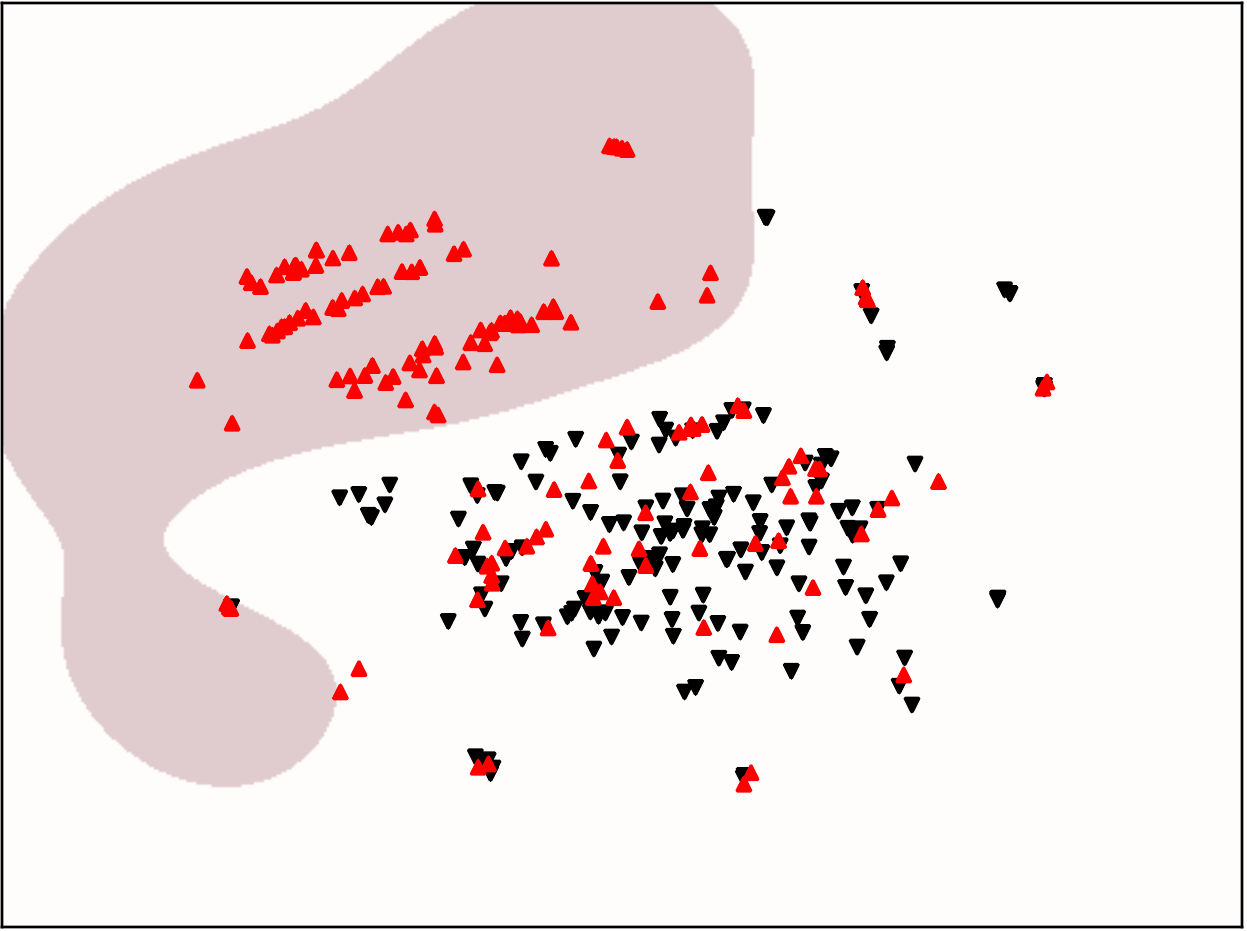}
        \caption{Software}
        \label{fig:ML_Clustering_Software2}
    \end{subfigure}
    \caption{The decision boundary of the risk-free and risky websites. A risky website is a website with potential malicious intention. Notice that this malicious behaviour can be characterized (i.e., determined) using a support vector machine.}
    \label{fig:ML_Clustering2}
\end{figure*}

\subsection{Risk Modeling}\label{sec:riskCharacterization}
The insights that we have provided thus far are intriguing, although we are left with a key question: how much of these insights can be generalized across sites of the same population and type for risk assessment? To answer this question, we report on our effort to identify  \textit{risky} websites using simple machine learning algorithms.

\BfPara{Risky Websites} A website in our analysis is considered risky if it is associated with any of the following:

\noindent(1)~\textit{Malicious Domain.} Websites that are associated with URLs responsible for malicious activities are considered risky~\cite{WestM14}.

\noindent(2)~\textit{Malicious Files.} Upon visiting a website, multiple scripts are executed on the host. As such, we consider any website with malicious files~\cite{KosbaMWTK14}, regardless of its \textit{VirusTotal} label, as a risky website. 

\noindent(3)~\textit{Blacklisted URLs.}  Blacklisting can occur due to (i) massive user reporting, or (ii) previous maliciousness by the website (e.g., scam attacks). As such, we consider all blacklisted websites as risky~\cite{WestM14}.

\noindent(4)~\textit{Vulnerable Websites.} Websites that are identified as vulnerable by \textit{Sucuri} are considered risky,  for the potential exploitability.

We note that the risk modeling is not limited to free content websites. We also consider any free content and premium website with one or more of the aforementioned aspects as a risky website and otherwise a risk-free website.

\BfPara{Website Features} To model the risks associated with each service, and as common in the relevant literature~\cite{MohaisenA13,WestM14,MohaisenA14a,Mohaisen15}, we leverage the aforementioned extracted features as a representation. In particular,~\autoref{tab:features_description} shows the superset of potential features that we use to represent each online service, including SSL certificate, page size, load time, TLD, and website content features. Additionally, we include three more features extracted using \textit{SecurityTrails}~\cite{securitytrails}: (i) the lifetime of a service IP address, (ii) whether a website is using or previously used Cloudflare as a Content Delivery Network (CDN), and (iii) whether a website is using or previously used Akamai Tech as a CDN. 

\noindent{\em Hold-out Data.} The data obtained by  \textit{VirusTotal} and \textit{Sucuri} (\#15--\#25) in \autoref{tab:features_description} is held out, and is only used to model validation. This allows us to utilize easy-to-obtain website quality metrics that do not require access to third-party information to model the website risk. We envision that our lightweight modeling, in contrast to third-party risk data, would be more practical, since the third-party labels are determined based on reporting and expensive analyses accumulated over a period of time. Solely relying on third-party tools, such as  \textit{VirusTotal} to identify risks would exclude a significant number of websites, including those newly created for free content. 

\BfPara{Risk Boundaries}
Considering the aforementioned features, we visualize the boundaries between \textit{risky} and \textit{risk-free} websites, shown in \autoref{fig:ML_Clustering2}.
In particular, we use the t-distributed stochastic neighbor embedding (t-SNE) visualization technique~\cite{van2008visualizing} to plot the features of the websites. Then, using a support vector machine, we estimate the risk boundaries, shown in the red-shaded area in~\autoref{fig:ML_Clustering2}. Based on the validation, we find that the riskiest websites are clustered together, as they share different website features. Our modeling is capable of identifying risky websites with an accuracy of 86.81\%, despite some limitations (e.g., potential false positives among our sampled websites).

\observation{
We address the need for lightweight risk modeling of free content websites using a representation of 17 generic and file-related features. Our modeling is shown to be effective, producing an accuracy of 86.81\%.}

\section{Conclusion and Future Work}\label{sec:conclusion}
Free content websites are an interesting element of the makeup of the web today, and their characteristics are not rigorously analyzed nor understood in contrast to other websites that offer the same content. This paper provides the first look into a comparative analysis of such websites across various domain- and content-level dimensions, as well as their risk profiles. Our curated datasets offer valuable resources for exploring this uncharted space, and our findings shed light on the fundamental differences between free content websites in contrast to premium websites. 

We believe that our analysis in this paper only ``scratches the surface'' of this important problem and calls for further explorations and actions. For instance, our domain- and content-level analyses have been only focused on easy-to-obtain metadata features and did not consider the in-depth features, e.g., linguistic, network topology information, regional information, deep content type, and organization attributes (e.g., in the case of SSL certificates; signing authorities, and hosting infrastructure). All of these dimensions could shed more light on the characteristics of such websites and constitute our future work. Finally, we notice that our analysis utilizes a single snapshot of those websites, and we did not consider the temporal dimension of their characteristics, which would be a very interesting yet challenging aspect to explore.

\end{document}